\documentclass[conference, 10 pt]{IEEEtran}
\IEEEoverridecommandlockouts

\usepackage{cite}
\usepackage{amsmath,amssymb,amsfonts,amsthm}
\usepackage[linesnumbered,ruled,vlined]{algorithm2e}
\usepackage{graphicx}
\usepackage{booktabs}
\usepackage{textcomp}
\usepackage{tabularx}
\newcolumntype{C}{>{\centering\arraybackslash}X}
\usepackage{hyperref}
\usepackage{cleveref}
\usepackage{orcidlink}
\usepackage{arydshln}
\usepackage{url}
\usepackage{soul}
\usepackage{comment}
\usepackage[in line]{enumitem}  
\usepackage{microtype}
\usepackage{subcaption} 
\usepackage[utf8]{inputenc} 
\usepackage{times}
\usepackage[T1]{fontenc}
\usepackage{tikz} 
\usetikzlibrary{positioning,arrows.meta,shapes.geometric,fit,patterns,spy}
\usepackage[nolist]{acronym}
\usepackage{optidef}
\usepackage{multirow}
\usepackage{pgfplots}
\usepackage{stfloats} 

\pgfplotsset{
    every axis/.append style = {
        font = \footnotesize,
        mark size = 2.75 pt,
        mark repeat = 16,
        line width = 1.5 pt,
        legend style={
            font=\footnotesize,
            at={(0.98,0.45)},
            anchor=north east,
            legend cell align=left,
            column sep=1ex, 
        },
        width= .9\linewidth, 
        height=.8\linewidth, 
    },
    compat=1.18
}

\newtheorem{theorem}{Theorem}
\newtheorem{corollary}{Corollary}

\newtheorem{definition}{Definition}
\newtheorem{remark}{Remark}
\newtheorem{lemma}{Lemma}
\renewcommand{\figurename}{Fig.}
\Crefname{equation}{Eq.\!}{Eqs.\!}
\Crefname{figure}{Fig.\!}{Figs.\!}
\Crefname{tabular}{Tab.\!}{Tabs.\!}
\Crefname{section}{Section\!}{Sections.\!}
\Crefname{appsec}{Appendix}{Appendices}

\newcommand{\indep}{\perp \!\!\! \perp}

\begin{document}

\def \NumOfSensingAntennas{n_{\textrm{s}}}
\def \RangeRV{R_{\textrm{s}}}
\def \RangeDet{r_{\textrm{s}}}
\def \deltars{\Delta\RangeDet}
\def \instRangeRV{R_{\textrm{s},i}}
\def \instRangeDet{r_{\textrm{s},i}}
\def \instRangeDetj{r_{\textrm{s},j,i}}
\def \RangeDetcomm{r_{\textrm{c}}}
\def \YCommRV{Y_{\textrm{c},i}}
\def \YCommRVsl{Y_{\textrm{c}}}
\def \YCommRVsldet{y_{\textrm{c}}}
\def \YSensRVsldet{\boldsymbol{y}_{\textrm{s}}}
\def \YSensRVsl{\boldsymbol{Y}_\textrm{s}}
\def \YSensingRV{\boldsymbol{Y}_{\textrm{s},i}}
\def \YSensingdetslj{y_{\textrm{s},j}}
\def \YSensingRVsl{\boldsymbol{Y}_{\textrm{s}}}
\def \FOV {\textrm{FOV}}
\def \ScalarCommCh{h_{\textrm{c}}}
\def \VectorSensingChsimple{\boldsymbol{h}_{\textrm{s}}}
\def \VectorSensingCh{\boldsymbol{h}_{\textrm{s},i}(r_{\textrm{s},j,i})}
\def \VectorSensingChDet{\boldsymbol{h}_\textrm{s}(\instRangeDet)}
\def \SensingChDetj{h_{\textrm{s},j}(\instRangeDet)}
\def \SensingChDetji{h_{\textrm{s},j,i}(\instRangeDetj)}
\def \TxsignalRV{X_{i}}
\def \TxsignalRVsl{X}
\def \TxsignalRVrealization{x_i}
\def \CommNoiseRV{Z_{\textrm{c},i}}
\def \CommNoiseRVsingleletter{Z_{\textrm{c}}}
\def \RCS{\rho}
\def \SensNoiseRV{\boldsymbol{Z}_{\textrm{s},i}}
\def \SensNoiseRVsingleletter{\boldsymbol{Z}_{\textrm{s}}}
\def \Rsset{\mathcal{R}_\textrm{s}}
\def \Xset{\mathcal{X}}
\def \Ysset{\mathcal{Y}_\textrm{s}}
\def \Ycset{\mathcal{Y}_c}
\def \Mset {\mathcal{M}}
\def \Xset {\mathcal{X}}
\def \Rshat {\hat{R_\textrm{s}}}
\def \Rshatdet {\hat{r_\textrm{s}}}
\def \Rshatdetinst{\hat{r_\textrm{s}}^\star_{i}}
\def \psic{\psi_\textrm{c}}
\def \phic{\phi_\textrm{c}}
\def \SensingNoiseVar{\sigma_{\textrm{s}}^2}
\def \CommNoiseVar{\sigma_{\textrm{c}}^2}
\def \psis{\psi_\textrm{s}}
\def \phis{\phi_\textrm{s}}

\begin{acronym}

\acro{5G-NR}{5G New Radio}
\acro{3GPP}{3rd Generation Partnership Project}
\acro{ACF}{autocorrelation function}
\acro{ACR}{autocorrelation receiver}
\acro{ADC}{analog-to-digital converter}
\acrodef{aic}[AIC]{Analog-to-Information Converter}     
\acro{AIC}[AIC]{Akaike information criterion}
\acro{aric}[ARIC]{asymmetric restricted isometry constant}
\acro{arip}[ARIP]{asymmetric restricted isometry property}
\acro{IM/DD}{intensity modulation direct detection}
\acro{OWC}{optical wireless communication}
\acro{V2X}{vehicle-to-everything}
\acro{ITS}{intelligent transportation system}
\acro{ARQ}{automatic repeat request}
\acro{AUB}{asymptotic union bound}
\acrodef{awgn}[AWGN]{Additive White Gaussian Noise}    
\acro{S-I}{signal-independent}
\acro{CFA}{\ac{CF} algorithm}
\acro{AWGN}{additive white Gaussian noise}
\acro{CRB}{Cramér-Rao bound}
\acro{BCRB}{Bayesian CRB}
\acro{BA}{Blahut–Arimoto}
\acro{BAA}{Blahut–Arimoto Algorithm}
\acro{ISAC}{integrated sensing and communication}
\acro{VCSEL}{vertical-cavity surface-emitting laser}
\acro{R-D}{rate-distortion}
\acro{R-CRB}{rate-\ac{CRB}}
\acro{C-D}{capacity-distortion}
\acro{EC-D}{enhanced capacity-distortion}
\acro{R-D}{rate-distortion}
\acro{TS}{time sharing}
\acro{APSK}[PSK]{asymmetric PSK}
\acro{SRC}{sensing response channel}
\acro{CC}{communication channel}
\acro{AoD}{angle of departure}
\acro{AoA}{angle of arrival}
\acro{waric}[AWRICs]{asymmetric weak restricted isometry constants}
\acro{warip}[AWRIP]{asymmetric weak restricted isometry property}
\acro{BCH}{Bose, Chaudhuri, and Hocquenghem}        
\acro{BCHC}[BCHSC]{BCH based source coding}
\acro{BEP}{bit error probability}
\acro{BFC}{block fading channel}
\acro{BG}[BG]{Bernoulli-Gaussian}
\acro{BGG}{Bernoulli-Generalized Gaussian}
\acro{BPAM}{binary pulse amplitude modulation}
\acro{BPDN}{Basis Pursuit Denoising}
\acro{BPPM}{binary pulse position modulation}
\acro{BPSK}{binary phase shift keying}
\acro{BPZF}{bandpass zonal filter}
\acro{BSC}{binary symmetric channels}              
\acro{BU}[BU]{Bernoulli-uniform}
\acro{BER}{bit error rate}
\acro{BS}{base station}
\acro{LS}{least squares}
\acro{SISO-COM and SIMO-SEN}{single-input single-output for communication and single-input multiple output for sensing}
\acro{CP}{Cyclic Prefix}
\acrodef{cdf}[CDF]{cumulative distribution function}   
\acro{CDF}{cumulative distribution function}
\acrodef{c.d.f.}[CDF]{cumulative distribution function}
\acro{CCDF}{complementary cumulative distribution function}
\acrodef{ccdf}[CCDF]{complementary CDF}               
\acrodef{c.c.d.f.}[CCDF]{complementary cumulative distribution function}
\acro{CD}{cooperative diversity}
\acro{MAP}{maximum a posteriori}
\acro{CDMA}{Code Division Multiple Access}
\acro{ch.f.}{characteristic function}
\acro{CIR}{channel impulse response}
\acro{cosamp}[CoSaMP]{compressive sampling matching pursuit}
\acro{CR}{cognitive radio}
\acro{cs}[CS]{compressed sensing}                   
\acrodef{cscapital}[CS]{Compressed sensing} 
\acrodef{CS}[CS]{compressed sensing}
\acro{CSI}{channel state information}
\acro{CCSDS}{consultative committee for space data systems}
\acro{Covid19}[COVID-19]{Coronavirus disease}
\acro{IB}{inner bound}

\acro{DAA}{detect and avoid}
\acro{DAB}{digital audio broadcasting}
\acro{DCT}{discrete cosine transform}
\acro{dft}[DFT]{discrete Fourier transform}
\acro{DR}{distortion-rate}
\acro{DS}{direct sequence}
\acro{DS-SS}{direct-sequence spread-spectrum}
\acro{DTR}{differential transmitted-reference}
\acro{DVB-H}{digital video broadcasting\,--\,handheld}
\acro{DVB-T}{digital video broadcasting\,--\,terrestrial}
\acro{DL}{downlink}
\acro{DSSS}{Direct Sequence Spread Spectrum}
\acro{DFT-s-OFDM}{Discrete Fourier Transform-spread-Orthogonal Frequency Division Multiplexing}
\acro{DAS}{distributed antenna system}
\acro{DNA}{Deoxyribonucleic Acid}

\acro{EC}{European Commission}
\acro{EED}[EED]{exact eigenvalues distribution}
\acro{EIRP}{Equivalent Isotropically Radiated Power}
\acro{ELP}{equivalent low-pass}
\acro{eMBB}{Enhanced Mobile Broadband}
\acro{EMF}{electric and magnetic fields}
\acro{EU}{European union}

\acro{FC}[FC]{fusion center}
\acro{FCC}{Federal Communications Commission}
\acro{FEC}{forward error correction}
\acro{FFT}{fast Fourier transform}
\acro{FH}{frequency-hopping}
\acro{FH-SS}{frequency-hopping spread-spectrum}
\acrodef{FS}{Frame synchronization}
\acro{FSsmall}[FS]{frame synchronization}  
\acro{FDMA}{Frequency Division Multiple Access}

\acro{GA}{Gaussian approximation}
\acro{GF}{Galois field }
\acro{GG}{Generalized-Gaussian}
\acro{GIC}[GIC]{generalized information criterion}
\acro{GLRT}{generalized likelihood ratio test}
\acro{GPS}{Global Positioning System}
\acro{GMSK}{Gaussian minimum shift keying}
\acro{GSMA}{Global System for Mobile communications Association}

\acro{HAP}{high altitude platform}

\acro{IDR}{information distortion-rate}
\acro{IFFT}{inverse fast Fourier transform}
\acro{iht}[IHT]{iterative hard thresholding}
\acro{i.i.d.}{independent, identically distributed}
\acro{IoT}{Internet of Things}                      
\acro{IR}{impulse radio}
\acro{lric}[LRIC]{lower restricted isometry constant}
\acro{lrict}[LRICt]{lower restricted isometry constant threshold}
\acro{ISI}{intersymbol interference}
\acro{ITU}{International Telecommunication Union}
\acro{ICNIRP}{International Commission on Non-Ionizing Radiation Protection}
\acro{IEEE}{Institute of Electrical and Electronics Engineers}
\acro{ICES}{IEEE international committee on electromagnetic safety}
\acro{IEC}{International Electrotechnical Commission}
\acro{IARC}{International Agency on Research on Cancer}
\acro{IS-95}{Interim Standard 95}

\acro{KKT}{Karush–Kuhn–Tucker}
\acro{LEO}{low earth orbit}
\acro{LF}{likelihood function}
\acro{LLF}{log-likelihood function}
\acro{LLR}{log-likelihood ratio}
\acro{LLRT}{log-likelihood ratio test}
\acro{LoS}{line-of-sight}
\acro{LRT}{likelihood ratio test}
\acro{wlric}[LWRIC]{lower weak restricted isometry constant}
\acro{wlrict}[LWRICt]{LWRIC threshold}
\acro{LPWAN}{low power wide area network}
\acro{LoRaWAN}{Low power long Range Wide Area Network}
\acro{NLOS}{non-line-of-sight}

\acro{MB}{multiband}
\acro{MC}{multicarrier}
\acro{MDS}{mixed distributed source}
\acro{MF}{matched filter}
\acro{m.g.f.}{moment generating function}
\acro{MI}{mutual information}
\acro{MIMO}{multiple-input multiple-output}
\acro{MISO}{multiple-input single-output}
\acrodef{maxs}[MJSO]{maximum joint support cardinality}                       
\acro{ML}[ML]{machine learning}
\acro{MLE}[MLE]{maximum likelihood estimator}
\acro{GD}{gradient descent}
\acro{MMSE}{minimum mean-square error}
\acro{LMMSE}{linear minimum mean-square error}
\acro{MSE}{mean-square error}
\acro{MMV}{multiple measurement vectors}
\acrodef{MOS}{model order selection}
\acro{M-PSK}[${M}$-PSK]{$M$-ary phase shift keying}                       
\acro{M-APSK}[${M}$-PSK]{$M$-ary asymmetric PSK} 

\acro{M-QAM}[$M$-QAM]{$M$-ary quadrature amplitude modulation}
\acro{MRC}{maximal ratio combiner}                  
\acro{maxs}[MSO]{maximum sparsity order}                                      
\acro{M2M}{machine to machine}                                                
\acro{MUI}{multi-user interference}
\acro{mMTC}{massive Machine Type Communications}      
\acro{mm-Wave}{millimeter-wave}
\acro{MP}{mean posterior}
\acro{MPE}{maximum permissible exposure}
\acro{MAC}{media access control}
\acro{NB}{narrowband}
\acro{NBI}{narrowband interference}
\acro{NLA}{nonlinear sparse approximation}
\acro{NLOS}{Non-Line of Sight}
\acro{NTIA}{National Telecommunications and Information Administration}
\acro{NTP}{National Toxicology Program}
\acro{NHS}{National Health Service}

\acro{OC}{optimum combining}                             
\acro{OC}{optimum combining}
\acro{ODE}{operational distortion-energy}
\acro{ODR}{operational distortion-rate}
\acro{OFDM}{orthogonal frequency-division multiplexing}
\acro{omp}[OMP]{orthogonal matching pursuit}
\acro{OSMP}[OSMP]{orthogonal subspace matching pursuit}
\acro{OQAM}{offset quadrature amplitude modulation}
\acro{OQPSK}{offset QPSK}
\acro{OFDMA}{Orthogonal Frequency-division Multiple Access}
\acro{OPEX}{Operating Expenditures}
\acro{OQPSK/PM}{OQPSK with phase modulation}

\acro{PAM}{pulse amplitude modulation}
\acro{PAR}{peak-to-average ratio}
\acrodef{pdf}[PDF]{probability density function}                      
\acro{PDF}{probability density function}
\acrodef{p.d.f.}[PDF]{probability distribution function}
\acro{PDP}{power dispersion profile}
\acro{PMF}{probability mass function}                         
\acro{ROST}{Repetition-One Shot Trade-Off}
\acrodef{p.m.f.}[PMF]{probability mass function}
\acro{PN}{pseudo-noise}
\acro{PPM}{pulse position modulation}
\acro{PRake}{Partial Rake}
\acro{PSD}{power spectral density}
\acro{PSEP}{pairwise synchronization error probability}
\acro{PSK}{phase shift keying}
\acro{PD}{power density}
\acro{8-PSK}[$8$-PSK]{$8$-phase shift keying}
\acro{RCS}{Reflectivity Coefficient}
\acro{AR-D-GT}{Achievable Rate-Distortion-Ground Truth}
\acro{FSK}{frequency shift keying}
\acro{CVAE}{Conditional Variational Auto-encoder}
\acro{FIM}{Fisher information matrix}

\acro{QAM}{Quadrature Amplitude Modulation}
\acro{QPSK}{quadrature phase shift keying}
\acro{OQPSK/PM}{OQPSK with phase modulator }

\acro{RS}[RS]{Reed-Solomon}
\acro{RSC}[RSSC]{RS based source coding}
\acro{r.v.}{random variable}                               
\acro{R.V.}{random vector}
\acro{RMS}{root mean square}
\acro{RFR}{radiofrequency radiation}
\acro{RIS}{Reconfigurable Intelligent Surface}
\acro{RNA}{RiboNucleic Acid}

\acro{SA}[SA-Music]{subspace-augmented MUSIC with OSMP}
\acro{SCBSES}[SCBSES]{Source Compression Based Syndrome Encoding Scheme}
\acro{SCM}{sample covariance matrix}
\acro{SEP}{symbol error probability}
\acro{SG}[SG]{sparse-land Gaussian model}
\acro{SIMO}{single-input multiple-output}
\acro{SINR}{signal-to-interference plus noise ratio}
\acro{SIR}{signal-to-interference ratio}
\acro{SISO}{single-input single-output}
\acro{SNR}[\textrm{SNR}]{signal-to-noise ratio} 
\acro{O-SNR}[\textrm{O-SNR}]{optical signal-to-noise ratio} 
\acro{SS}{spread spectrum}

\acro{TH}{time-hopping}
\acro{ToA}{time-of-arrival}
\acro{TR}{transmitted-reference}
\acro{TW}{Tracy-Widom}
\acro{TWDT}{TW Distribution Tail}
\acro{TCM}{trellis coded modulation}
\acro{TDD}{time-division duplexing}
\acro{TDMA}{time division multiple access}
\acro{DRT}{deterministic-random tradeoff}
\acro{UAV}{unmanned aerial vehicle}
\acro{uric}[URIC]{upper restricted isometry constant}
\acro{urict}[URICt]{upper restricted isometry constant threshold}
\acro{SC}[S\&C]{sensing and communication}
\acro{O-ISAC}{optical ISAC}
\acro{UWB}{ultrawide band} 
\acro{URLLC}{Ultra Reliable Low Latency Communications}
\acro{FOV}{field of view}                      
\acro{UE}{user equipment}
\acro{UL}{uplink}
\acro{ULA}{uniform linear array}
\acro{P2P}{point-to-point}
\acro{SDMC-DF}{state-dependent memoryless channel with delayed feedback}
\acro{Com. Rx}{communication receiver}
\acro{Tx}{transmitter}
\acro{Sens. Rx}{sensing receiver}
\acro{WLAN}{wireless local area network}
\acro{WMAN}{wireless metropolitan area network}
\acro{WPAN}{wireless personal area network}
\acro{WSN}{wireless sensor network}                        
\acro{Wi-Fi}{wireless fidelity}
\acro{CF}{closed-form}
\acro{OB}{outer bound}
\acro{UB}{upper bound}
\acro{LB}{lower bound}
\acro{CLT}{central limit theorem}
\acro{LiSAC}{Lighting, Sensing, and Communication}
\acro{VLC}{visible light communication}
\acro{VPN}{virtual private network} 
\acro{RF}{radio frequency}
\acro{FSO}{free space optical}
\acro{Com. Opt.}{communication optimal}
\acro{Sens. Opt.}{sensing optimal}
\acro{w.r.t.}{with respect to}
\acro{LiDAR}{Light Detection and Ranging}
\acro{GSM}{Global System for Mobile Communications}
\acro{2G}{second-generation cellular network}
\acro{3G}{third-generation cellular network}
\acro{4G}{fourth-generation cellular network}
\acro{5G}{5th-generation cellular network}

\acro{QoS}{quality of service}
\end{acronym}

\newcommand\mycommfont[1]{\footnotesize\ttfamily\textcolor{blue}{#1}}
\SetCommentSty{mycommfont}
\SetKwFunction{BAAF}{$[P_i, D_i, \mathcal{I}_i, P_X^\star] := \ac{BAA}-F$}
\SetKwFunction{CFAF}{$[P_i, D_i, P_\TxsignalRVsl, \mathcal{I}_i] := \ac{CFA}-F$}
\SetKwInput{KwInput}{Input}                
\SetKwInput{KwOutput}{Output}              
\SetKwInput{KwInit}{Initialization}
\SetKwProg{Fn}{Function}{:}{}

\title{Optical ISAC: Fundamental Performance Limits and Transceiver Design\\
}
\author{\IEEEauthorblockN{Alireza Ghazavi Khorasgani\orcidlink{0009-0001-9141-7260}, Mahtab Mirmohseni\orcidlink{0000-0002-5247-5820}, Ahmed Elzanaty\orcidlink{0000-0001-8854-8369}
}
\IEEEauthorblockA{\textit{5/6GIC, Institute for Communication Systems (ICS), \textit{University of Surrey}, Guildford, United Kingdom} \\
\{\href{mailto:a.ghazavi@surrey.ac.uk}{a.ghazavi}, \href{mailto:m.mirmohseni@surrey.ac.uk}{m.mirmohseni}, \href{mailto:a.elzanaty@surrey.ac.uk}{a.elzanaty}\}@surrey.ac.uk
}}

\maketitle

\begin{abstract}
This paper characterises the optimal capacity-distortion (C-D) tradeoff in an optical point-to-point system with single-input single-output (SISO) for communication and single-input multiple-output (SIMO) for sensing within an integrated sensing and communication (ISAC) framework. We consider the optimal rate-distortion (R-D) region and explore several inner (IB) and outer bounds (OB). We introduce practical, asymptotically optimal maximum a posteriori (MAP) and maximum likelihood estimators (MLE) for target distance, addressing nonlinear measurement-to-state relationships and non-conjugate priors. As the number of sensing antennas increases, these estimators converge to the Bayesian Cramér-Rao bound (BCRB). We also establish that the achievable rate-Cramér-Rao bound (R-CRB) is an OB for the optimal C-D region, valid for unbiased estimators and asymptotically large numbers of receive antennas. To clarify that the input distribution determines the tradeoff across the Pareto boundary of the C-D region, we propose two algorithms: \textit{i}) an iterative Blahut-Arimoto algorithm (BAA)-type method, and \textit{ii}) a memory-efficient closed-form (CF) approach. The CF approach includes an optimal distribution for high optical signal-to-noise ratio (O-SNR) conditions. Additionally, we adapt and refine the deterministic-random tradeoff (DRT) to this optical ISAC context.
\end{abstract}

\section{Introduction}

Future wireless networks are integrating advanced \ac{SC} technologies, critical for applications such as intelligent transportation systems and smart cities. \Ac{ISAC} systems reflect this synergy, where \ac{SC} functionalities share hardware, spectrum, and signaling resources \cite{SeventyYears,tishchenkoDualFunctionalMm2024}. \Ac{O-ISAC} is a promising alternative to \ac{RF} \ac{ISAC}, especially in \ac{FSO} systems, which leverage the large bandwidth of optical signals for high-speed communication and high-resolution sensing \cite{LISAC}. In transportation, \ac{O-ISAC} provides low latency, high data rates, and access to unlicensed spectrum, significantly improving \ac{V2X} communication, traffic safety \cite{An2023}, and disaster management \cite{Qazavi}. Unlike \ac{RF}, \ac{O-ISAC} experiences minimal interference in dense traffic due to laser beam directivity, reducing self-interference and enhancing security via \ac{LoS} links \cite{An2023}. \ac{FSO} is especially promising for \ac{V2X} \cite{An2023}, where straight-line vehicle movement simplifies \ac{AoA} assumptions. Vehicles can also use \ac{LiDAR} to gather state info, reducing direct communication (sensing-assisted communication and vice versa). \ac{LiDAR}-based \ac{ISAC} supports omnidirectional communication, enabling flexible links beyond headlights' range for cooperative driving.

Previous works have focused on \ac{RF} \ac{ISAC} systems \cite{mehrasajournal,xiong2023torch,HuaJournal,Marziyeh,Vehicular,RenJournal}. In \cite{mehrasajournal}, the optimal \ac{C-D} for single-antenna \ac{RF} \ac{ISAC} systems was examined, with the optimal estimator simplifying to the \ac{LMMSE} estimator under specific conditions, such as Gaussian priors. However, real-world scenarios often involve nonlinear functions of \ac{SRC} and non-conjugate priors, complicating optimal estimator computation \cite{alma991028260009704706}. Significant research has addressed the \ac{R-CRB} tradeoff in \ac{RF} channels, particularly concerning parameters like \ac{AoD} and \ac{AoA} \cite{xiong2023torch,HuaJournal,RenJournal,Marziyeh}. Despite these efforts, gaps remain in practical estimators, transceiver design, and optimal \ac{C-D} regions. Most studies have concentrated on Gaussian signaling, which may not fully exploit the potential benefits for \ac{ISAC}. To enhance observations, one proposed solution is to record multiple feedbacks across several channel uses with block-wise \ac{i.i.d.} states (block length $T$). While this improves sensing performance, it proportionally degrades communication performance at a rate of $T^{-1}$, especially for large $T$ \cite[Eq. 36]{xiong2023torch}. While existing works focus on \ac{RF} signals, which differ from optical systems due to their positive, real nature, \ac{O-ISAC} systems, to the best of our knowledge, have not been explored regarding information-theoretical limits.

This paper makes several key contributions: \textit{i}) We characterise the optimal Pareto boundary of the \ac{R-D} and \ac{C-D} regions for \ac{O-ISAC} systems, which leverages multiple antennas to enhance both \ac{SC}, focusing particularly on target distance estimation with nonlinear \ac{SRC} relationships and non-conjugate priors. \textit{ii}) We adapt and refine the \ac{DRT} \cite{xiong2023torch} for \ac{O-ISAC} and general estimators, introducing practical, asymptotically optimal estimators. We analyze the performance of our proposed \ac{MAP} and \ac{MLE} estimators, demonstrating their convergence to the \ac{CRB} as the number of sensing antennas increases. \textit{iii}) We demonstrate that, in asymptotic scenarios, the achievable \ac{R-CRB} serves as an \ac{OB}, while the \ac{MAP}, \ac{MLE}, and any unbiased estimator function as an \ac{IB} for the optimal \ac{C-D} region. \textit{iv}) We propose two algorithms to determine the optimal input distribution for the Pareto boundary of the \ac{C-D} region, validate these algorithms against the endpoints, and characterise the optimal \ac{O-ISAC} input distribution for high \ac{O-SNR}.

\textbf{Notation:} Sets are denoted by calligraphic letters (e.g., $\mathcal{X}$), with cardinality $\lvert \mathcal{X} \rvert$. Real numbers are $\mathbb{R}$; nonnegative reals are $\mathbb{R}_0^+$. Random variables are uppercase (e.g., $\TxsignalRVsl$), and realisations are lowercase (e.g., $x$). Vectors are boldfaced (e.g., $\YSensRVsl$). Key symbols include $\sim$ (distribution), $\indep$ (independence), and $\overset{a}{\sim}$ (asymptotic distribution). Functions/operators: $\mathcal{N}(\mu, \sigma^2)$ (Gaussian), $\mathcal{H}(\cdot)$ (entropy), $\mathcal{I}(\cdot)$ (\ac{MI}), $\mathbb{E}_X[\cdot]$ (expectation), $| \cdot |$ (absolute value), and $\| \cdot \|$ ($\ell_2$ norm).

\section{System Model}
\begin{figure}[t]
\centering
\includegraphics[width=0.85\linewidth]{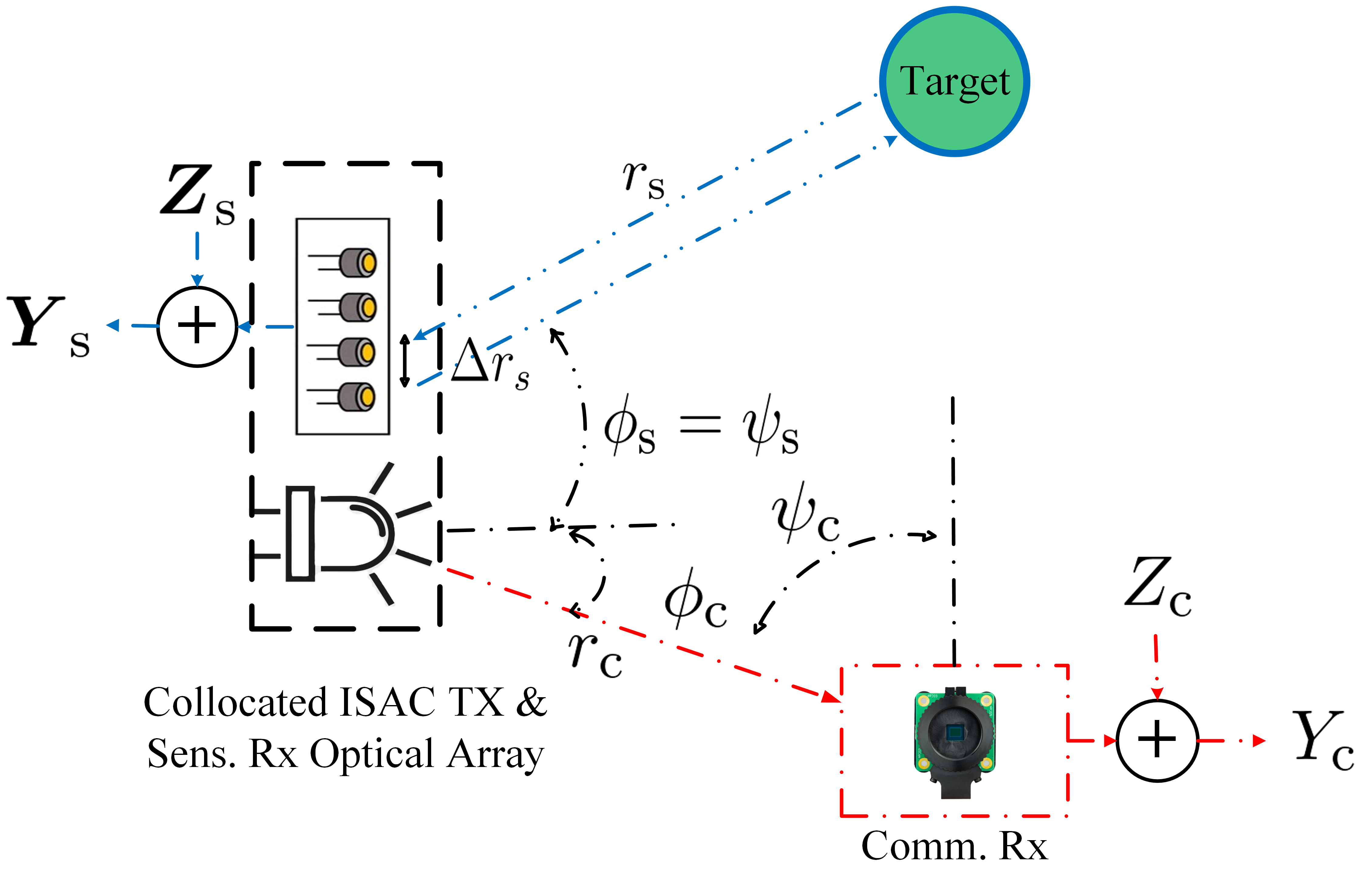} 
\caption{\ac{O-ISAC} system with memoryless channels.}
\label{fig:isac_models}
\end{figure}

We consider a point-to-point \ac{O-ISAC} system as illustrated in \Cref{fig:isac_models}. This system comprises a single-antenna \ac{Tx},
an $\NumOfSensingAntennas$-antenna monostatic \ac{Sens. Rx} that is collocated with the \ac{Tx}, a single-antenna \ac{Com. Rx}, and a point-wise target. This configuration is typical for \ac{LiDAR} evaluations \cite{straighline}. In this setup, data is transmitted to the \ac{Com. Rx} while simultaneously estimating the target distance $\RangeRV \in \mathbb{R}_0^{+}$, with the realization denoted as $\RangeDet$. The distance estimation is based on echoes received at the \ac{Sens. Rx}. Additionally, we use a state-dependent \ac{FSO} \ac{ISAC} channel with \ac{IM/DD}, which is detected by both receivers \cite{ahmad,ahmed2}. The input signal is constrained to be nonnegative due to the optical nature of the system, and its average power must satisfy a given optical power budget.

\textbf{\ac{SC} Models:} The received signal at \ac{Com. Rx} during the $i$-th channel use is:
\begin{equation}
    \YCommRV = \ScalarCommCh \TxsignalRV + \CommNoiseRV,
    \label{eq:transmission_model}
\end{equation}
where $\ScalarCommCh \in \mathbb{R}$ is the \ac{LoS} channel, $\TxsignalRV$ is the transmitted signal, and $\CommNoiseRV \sim \mathcal{N}(0, \CommNoiseVar)$ represents \ac{i.i.d.} \ac{AWGN}.
\begin{equation}
    \ScalarCommCh = \frac{A}{\RangeDetcomm^2} R_\textrm{0}(\phic) T_\textrm{s}(\psic) g(\psic) \cos{\psic} \cdot \mathbf{1}_{[0, \FOV]}(\psic),
\end{equation}
where $\RangeDetcomm$ is the distance between \ac{Tx} and \ac{Com. Rx}, $\phic$ and $\psic$ are angles relative to \ac{Tx} and \ac{Com. Rx}, $T_\textrm{s}(\cdot)$ and $g(\cdot)$ are the concentrator gain for \ac{Tx} and \ac{Com. Rx} respectively, $A$ is the effective area, and $\FOV$ is the \ac{FOV}. The \ac{Tx} radiant intensity gain is $R_\textrm{0}(\phi) = \frac{(m+1)}{2\pi} \cos^m{\phi}$, where $m = -\frac{\ln{2}}{\ln{(\cos{\Phi_{1/2}})}}$ \cite{ahmad}.

The echo signal at the $i$-th \ac{Sens. Rx} channel is:
\begin{equation}
    \YSensingRV = \VectorSensingCh \TxsignalRV + \SensNoiseRV,
    \label{eq:sensing_model}
\end{equation}
where $\VectorSensingCh \in \mathbb{R}^{\NumOfSensingAntennas \times 1}$ represents the target response coefficient, which depends on $\instRangeDetj$, the range (distance) of the point-wise target from \ac{Tx}. The sensing noise $\SensNoiseRV$ is modeled as $\SensNoiseRV \sim \mathcal{N}(0, \sigma_{\text{S}}^2 \boldsymbol{I}_{\NumOfSensingAntennas})$. The target response matrix $\SensingChDetji$ is given by \cite[Eq. 4]{vlc_sensing_survey}:
\begin{equation}
    \SensingChDetji = 
    \frac{\RCS}{\instRangeDetj^4} R_\text{0}(0) T_\text{s}(\psis) g(\psis) \cos{\psis} \cdot \mathbf{1}_{[0, \FOV]}(\phis),
\end{equation}
where $ \RCS = A^2 R_0(\phis) T_s(0) g(0) R_0(0) T_s(\psis) g(\psis) \cos{\phis} $ denotes the reflectivity coefficient, assumed to be a constant deterministic value without reflection-induced noise \cite{richards2010principles}. This assumption simplifies the analysis and offers baseline insights into system performance.
With \ac{Sens. Rx} antennas in a \ac{ULA} and the target moving straight, both sensing angles ($\psis$ and $\phis$) are equal. Assuming $\RangeDet \gg \deltars$ ($\deltars$ being the inter-antenna distance), we approximate $\instRangeDetj \approx \instRangeDet$ for all $i,j$, justifying the straight-line assumption as phase shifts are immeasurable in \ac{IM/DD} \cite{ahmad}. A uniform $8 \times 8$ array with $11.2 \, \mu\text{m}$ spacing achieves an \ac{FOV} of $8^\circ$ \cite{Fatemi:18}.

\textbf{Code Definition:} A $(2^{nR}, n)$ code for \ac{SDMC-DF} includes several components. First, there is a discrete message set $\Mset$ with $\lvert M \rvert = 2^{nR}$. Second, encoding functions $\phi_i : \Mset \times \Ysset^{i-1} \mapsto \Xset$ are defined for $i = 1, 2, \dots, n$. Third, a decoding function $f : \Rsset^n \times \Ycset^n \mapsto \Mset$ is provided. Fourth, a state estimator $h : \Xset \times \Ysset^n \mapsto \Rshat^n$ is included, with $\Rshat$ as the reconstruction alphabet. The random message $M$ is uniformly distributed over $\Mset$ for a given code. Inputs are generated as $\TxsignalRV = \varphi_i(W, \YSensingRVsl^{i-1})$ for $i = 1, \dots, n$. The channel outputs $\YCommRV$ and $\YSensingRV$ at time $i$ depend on the state $\instRangeRV$ and the input $\TxsignalRV$. These dependencies are governed by the transition laws $P_{\YCommRVsl \mid \TxsignalRVsl, \RangeRV}(\cdot \mid \TxsignalRVrealization, \instRangeDet)$ and $P_{\YSensingRVsl \mid \TxsignalRVsl, \RangeRV}(\cdot \mid \TxsignalRVrealization, \instRangeDet)$, as given in \eqref{eq:sensing_model} and \eqref{eq:transmission_model}. Let $\Rshat^n \triangleq (\hat{R}_{\textrm{s},1}, \dots, \hat{R}_{\textrm{s},n}) = h(X^n, \YCommRVsl^n)$ denote state estimate at \ac{Tx}, and $\hat{W} = g(\RangeRV^n, \YCommRVsl^n)$ the decoded message at \ac{Com. Rx}. The expected average per-block distortion measures the quality of state estimation:
$\Delta^{(n)} \triangleq \mathbb{E}[d(\RangeRV^n, \Rshat^n)] = \frac{1}{n} \sum_{i=1}^{n} \mathbb{E}[d(\instRangeRV, \hat{\instRangeRV})]$,
where $d : \Rsset \times \Rshat \to \mathbb{R}_0^+$ is a bounded distortion function with $\max_{(\RangeDet, \Rshatdet) \in \Rsset \times \hat{\mathcal{R}}_s} d(\RangeDet, \Rshatdet) < \infty$. In practical optical systems, $\TxsignalRVsl$ is proportional to optical intensity and thus nonnegative: $X \in \mathbb{R}_0^+$. Then,
$\mathbb{E}[|X^n|] = \frac{1}{n} \sum_{i=1}^{n} \mathbb{E}[|\TxsignalRV|]$\footnote{The monostatic \ac{Sens. Rx}, collocated with \ac{Tx}, knows $X^n$ \cite{xiong2023torch}. It estimates $\RangeRV$ from $\YSensingRVsl^n$, while \ac{Com. Rx} decodes $M$ from $\YCommRVsl^n$.}.
\begin{definition}{(\ac{C-D} Region)}
A \ac{C-D} tuple $(C, D)$ is achievable with power budget $P$ if there exists a sequence of $(2^{nR}, n)$ codes that satisfies:
\begin{align}
    \Delta^{(n)} \leq D, \quad \mathbb{E}[\lvert X^n \rvert] \leq P, \quad \text{P}_e^{(n)} \to 0.
\end{align}
Here, $\text{P}_e^{(n)} \triangleq \frac{1}{2^{nR}} \sum_{i=1}^{2^{nR}} \mathbb{P}\{\mathsf{\hat{M}} \neq i \mid M = i\}$ and $d(\RangeDet, \Rshatdet) = (\RangeDet - \Rshatdet)^2$ is the squared error distortion. \ac{C-D} region $C_P(D)$ for power budget $P$ is defined as $C_P(D) = \sup\{R \mid (R, D) \text{ is achievable with } P\}$.
\end{definition}

In the next section, we characterise the \ac{C-D} region for the \ac{O-ISAC} system. We first describe the optimal estimator $h$, which operates on a single input symbol with $\NumOfSensingAntennas$-fold feedback, estimating $\hat{\RangeRV}_{i}$ based solely on $\TxsignalRV$ and $\{Y_{\textrm{s},j,i}\}_{j=1}^{\NumOfSensingAntennas}$, excluding other feedback signals $\{Y_{\textrm{s},j,i^\prime}\}_{i^\prime \neq i}$. \Cref{lemma:est} shows that the optimal estimator relies solely on the current feedback $\YSensingRV$ from the memoryless \ac{SRC}, making the sensing cost a function of the input signal, $c(x)$ \cite[Lemma 1]{mehrasajournal}.
\begin{lemma}
\label{lemma:est}
The deterministic \ac{MMSE} estimator $\Rshatdet$, which minimises the expected distortion, depends on $x$ and $\YSensRVsldet \triangleq \text{vec}\{\YSensingdetslj\}_{j=1}^{\NumOfSensingAntennas}$ and is given by
\begin{equation}
\Rshatdet = \mathbb{E}_{\RangeRV}[\RangeRV \mid X=x, \YSensingRVsl = \YSensRVsldet].
\label{eq:optimal est.}
\end{equation}
The distortion $\Delta^{(n)}$ is minimised by the estimator, regardless of the encoding and decoding functions:
\begin{equation}
h^{\ast}(x^{n}, \YSensRVsldet^{n}) \triangleq (\Rshatdet^{\ast}(x_{1}, \boldsymbol{y}_{s,1}), \cdots, \hat{r}_s^{\ast}(x_{n}, \boldsymbol{y}_{s,n})).
\label{eq:jmehrasa:7}
\end{equation}
The estimation cost for each input symbol $x \in \Xset$ is
\begin{equation}
c(x) = \int P_{\RangeRV}\int P_{\YSensingRVsl \mid \TxsignalRVsl, \RangeRV}\times(\RangeDet - \Rshatdet(x, \YSensRVsldet))^2 d\YSensRVsldet d\RangeDet.
    \label{eq:c_x2} 
\end{equation}
\begin{proof}
See Appendix \ref{proof:lemma:est}.
\end{proof}
\end{lemma}
\section{Main Result}
In this section, we characterise the \ac{R-D} region through:
\begin{enumerate*}[label=(\roman*)]
    \item A Time-sharing-based scheme for \ac{Com. Opt.} and \ac{Sens. Opt.} modes,
    \item A \ac{CFA} for high \ac{O-SNR} scenarios (for \ac{Com. Rx}),
    \item A \ac{BAA}-type algorithm for general \ac{S-I} noise channels.
\end{enumerate*}
Additionally, we propose \ac{R-D} regions based on:
\begin{enumerate*}[label=(\roman*)]
    \item The \ac{MAP} estimator (\Cref{mapcost}),
    \item The \ac{MLE} (\Cref{mlcost}),
    \item \ac{BCRB}-based results yielding \ac{OB} (\Cref{sec:bcrb}).
\end{enumerate*}
To determine the \ac{C-D} region for the joint distribution $P_X P_{\RangeRV} P_{\YCommRVsl \mid X \RangeRV} P_{\YCommRVsl \mid X \RangeRV} P_{\RangeRV \mid \TxsignalRVsl, \YSensRVsldet}$, we leverage the results from \cite{mehrasajournal} by setting the channel state to $ \RangeRV $.

\begin{maxi!}[l]
    {P_X}{\mathcal{I}(X; \YCommRVsl \mid \RangeRV),}
    {\label{opt1}}{}
    \addConstraint{\int_{x \in \Xset} x P_X(x)}{\leq P \label{opt1a}}
    \addConstraint{\int_{x \in \Xset} c(x) P_X(x)}{\leq D \label{opt1b}}
    \addConstraint{\int_{x \in \Xset} P_X(x)}{= 1 \label{opt1c}}
    \addConstraint{x}{>0,\quad \forall x \in \mathcal{X}. \label{opt1d}}
\end{maxi!}
\textbf{1) Time-Sharing Scheme:}
This scheme involves time sharing between the following two modes:

\textbf{1-1) \ac{Com. Opt.}:} 
"Com. opt." denotes communication optimisation. Ignoring the distortion constraint, \eqref{opt1b} simplifies to the channel capacity $\mathcal{C}(\ScalarCommCh, P)$. Although the exact capacity formula is unknown, it is bounded by its \ac{UB} \cite[Theorem 8]{Moser} and \ac{LB} \cite[Example 12.2.5]{cover}.

\begin{align}
    \begin{split}
        \mathcal{C}_{\textrm{\ac{LB}}}(\ScalarCommCh, P) &= \frac{\frac{1}{2} \ln (\ScalarCommCh P) - \sqrt{\frac{\pi \sigma^2}{2 \ScalarCommCh P}} + \frac{1}{2} \ln \big(1 + \frac{2}{\ScalarCommCh P}\big)}{\ln 2} \\
        &\quad + \frac{\sqrt{\ScalarCommCh P (2 + \ScalarCommCh P)} - \ScalarCommCh P - 1}{\ln 2},  
    \end{split} \\
    \mathcal{C}_{\textrm{\ac{UB}}}(\ScalarCommCh, P) &= \frac{1 - \ln \big(\frac{1}{P}\big) + \ln (\ScalarCommCh)}{\ln 2}+o_{P}(1).
\end{align}
Here, $o_{P}(1)$ approaches zero as $P \to \infty$.

\noindent
\textbf{1-2) \ac{Sens. Opt.}:}
In this mode, \eqref{opt1} simplifies to identifying the input distribution that minimises average sensing distortion:
\begin{maxi!}[l]
    {P_X}{\int_{x \in \Xset} c(x) P_X(x),}
    {\label{opt3}}{}
    \addConstraint{\eqref{opt1a},\eqref{opt1c},\eqref{opt1d}}{\label{opt3a}.}
\end{maxi!}
\begin{lemma}
\label{lemm:sop}
The optimal solution to \eqref{opt3} is $P_X^{\text{\ac{Sens. Opt.}}}(x) = \delta(x - x^\star)$, where $x^\star \triangleq \arg\min_{x \leq P} c(x)$. This yields zero \ac{MI} and minimum distortion $D_{\text{min}} = c(x^\star)$. 
    \begin{proof}
See Appendix \ref{proof:sop}.
\end{proof}
\end{lemma}
\noindent \textbf{2) Optimal \ac{R-D} Region: \ac{CF} for High \ac{O-SNR} Regime:} In high \ac{O-SNR} regime ($\textrm{\ac{O-SNR}} \triangleq \frac{\mathbb{E}[X]}{\CommNoiseVar} \to \infty$), where $\mathcal{H}(\YSensingRVsl \mid X)$ is independent of $P_X$, \eqref{opt1} simplifies due to additive noise.
\begin{maxi!}[l]
    {P_X}{\mathcal{H}(X \mid \RangeRV) \overset{(a)}{=}\mathcal{H}(X),}
    {\label{opt2}}{}
\addConstraint{\eqref{opt1a}, \eqref{opt1b}, \eqref{opt1c},\eqref{opt1d}.}{\label{opt2a}}
\end{maxi!}
where (a) follows from $X \indep \RangeRV$ as per \cite[Theorem 1]{mehrasajournal}. Since \eqref{opt2} is a convex problem with concave entropy $\mathcal{H}(X)$ and affine constraints, it can be solved using the \ac{KKT} method \cite{Boyd_Vandenberghe_2004}.
\begin{lemma}
\label{lemma:kkt}
The solution to \eqref{opt2} is an exponential family \ac{PDF} given by
\begin{equation}
    p_{X}(x) = \exp{(1 - \eta_1 - \eta_2 x - \eta_3 c(x))},
    \label{eq:cf}
\end{equation}
where $\eta_1$ is the normalisation constant, while $\eta_2 \geq 0$ and $\eta_3 \geq 0$ are the dual variables for the power budget and sensing constraint, respectively.
\begin{proof}
The result follows from entropy definitions and the Lagrangian derivative. Details are omitted for brevity.
\end{proof}
\end{lemma}
\noindent \textbf{3) Optimal \ac{R-D} Region: \ac{BAA}-Type for General Cases:}
To solve \eqref{opt1} and derive the optimal \ac{C-D} region, we use the \ac{BAA} method \cite[Section VI]{1054855} for the general case and \cref{lemma:kkt} for the high-\ac{O-SNR} regime. We introduce two non-negative penalty factors, $\eta_2$ and $\eta_3$. For each fixed $\eta_3$ (representing a given distortion level), the optimal $\eta_2$ is determined by complementary slackness. Specifically, if $\eta_2 = 0$ satisfies the power budget constraint, then $\eta_2^\star = 0$; otherwise, $\eta_2^\star > 0$, and we adjust $\eta_2$ using gradient descent \cite{Boyd_Vandenberghe_2004} to satisfy the power budget constraint with equality.
The detailed algorithm is provided in Appendix \ref{app:numerical_methods}.
\begin{remark}
    In \ac{Com. Opt.} mode, we can set $\eta_3$ in \eqref{eq:cf} to zero, which results in an exponential distribution. This confirms the result presented in \cite{Moser}.
\end{remark}
To compute \eqref{eq:optimal est.}, we need $P_{\RangeRV \mid \TxsignalRVsl, \YSensRVsldet}(\RangeDet \mid x, \YSensRVsldet)$:
\begin{equation}
P_{\RangeRV \mid \TxsignalRVsl, \YSensRVsldet}(\RangeDet \mid x, \YSensRVsldet) = \frac{P_{\YSensRVsldet \mid \TxsignalRVsl, \RangeRV} P_{\RangeRV}}{\int_{\RangeDet \in \Rsset} P_{\YSensRVsldet \mid \TxsignalRVsl, \RangeRV} P_{\RangeRV}}.
    \label{eq:posterior}
\end{equation}
However, computing \eqref{eq:posterior} is generally intractable due to the complexity of the marginal distribution \cite{alma991028260009704706}.

\begin{lemma}
Let $\hat{\RangeDet}_{MAP}$ and $\hat{\RangeDet}_{MP}$ denote \ac{MAP} estimate $\arg \max_{\RangeDet\geq 0} P_{\RangeRV \mid \TxsignalRVsl, \YSensRVsldet}(\RangeDet \mid x, \YSensRVsldet)$ and the \ac{MP} estimate $\mathbb{E}_{\RangeDet}[P_{\RangeRV \mid \TxsignalRVsl, \YSensRVsldet}(\RangeDet \mid x, \YSensRVsldet)]$, respectively. Then, as $n_s \to \infty$, $\hat{\RangeDet}_{MAP} \to \hat{\RangeDet}_{MP}$ in probability and $P_{\RangeRV \mid \YSensRVsldet, X}$ has a Gaussian \ac{PDF}. Specifically,
$\Rshatdet \overset{a}{\sim} \mathcal{N}(\RangeDet, I^{-1}(\RangeDet))$.
\begin{proof}
By the Bernstein–von Mises theorem \cite{vaartAsymptoticStatistics1998} and \cite[Theorem 11.3]{kay}, for a large sample size $n_s$, $P_{\RangeRV \mid \TxsignalRVsl, \YSensRVsldet}(\RangeDet \mid x, \YSensRVsldet)$ is asymptotically normal with mean $\hat{\RangeDet}_{MP}$ and variance $\Sigma_n$, where $\Sigma_n$ is the inverse of the \ac{FIM}. 
Therefore, \ac{MAP} estimate $\hat{\RangeDet}_{MAP}$, which is the mode of the \eqref{eq:posterior}, converges to the mean of the \eqref{eq:posterior} $\hat{\RangeDet}_{MP}$.
\end{proof}
\end{lemma}
\subsection{\texorpdfstring{\ac{MAP}-Based Achievable \ac{ISAC} \ac{C-D} Region}{MAP-Based Achievable ISAC C-D Region}}
\label{mapcost}
\begin{theorem}
\label{mlth}
\ac{MAP} estimator ($\arg \max_{\RangeDet\geq0} P_{\RangeRV \mid \TxsignalRVsl, \YSensRVsldet}(\RangeDet \mid x, \YSensRVsldet)$) is $\Rshatdet \in \Rsset$ that satisfies:
\begin{equation}
\frac{\lambda\SensingNoiseVar}{\NumOfSensingAntennas} \Rshatdet^{9} + 4\RCS x \left(\frac{1}{\NumOfSensingAntennas}\sum_{j=1}^{\NumOfSensingAntennas}\YSensingdetslj\right) \Rshatdet^{4} -  4\RCS^2 x^2 = 0, \quad \Rshatdet \geq 0.
\label{eq:ml}
\end{equation}
\begin{proof}
Setting the curvature of the logarithm of \eqref{eq:posterior} with respect to $\RangeDet$ confirms the result.
\end{proof}
\end{theorem}

\Cref{eq:ml} can be solved numerically for any $x \in \Xset$ and $\YSensRVsldet \in \Ysset$ using methods such as Newton-Raphson.
Deriving an analytical \ac{PDF} for \ac{MAP} estimate is generally infeasible \cite{kay}. Instead, we use computer simulations for performance assessment, as detailed in \Cref{sec:num}.
\subsection{\texorpdfstring{\ac{MLE}-Based Achievable \ac{ISAC} \ac{C-D} Region}{MLE-Based ISAC C-D Region}}
\label{mlcost}

\begin{lemma}
Define \ac{MLE} as \[\hat{\RangeDet}_{MLE} = \arg \max_{\RangeDet \geq 0} P_{\YSensRVsldet \mid \TxsignalRVsl, \RangeRV}(\YSensRVsldet \mid x, \RangeDet)\]
As $\NumOfSensingAntennas \to \infty$, \ac{MAP} estimate approaches \ac{MLE}.
\label{maptoml}
\begin{proof}
As $\NumOfSensingAntennas \to \infty$, the logarithm of \eqref{eq:posterior} is dominated by the sum term $\sum_{s=1}^{\NumOfSensingAntennas} \big( \YSensRVsldet - \frac{\RCS x}{\RangeDet^4} \big)^2$,
while the term $-\lambda \RangeDet$ becomes negligible. Thus, the logarithm of \eqref{eq:posterior}
approximates the $\log(P_{\YSensRVsldet \mid \TxsignalRVsl, \RangeRV}(\YSensRVsldet \mid x, \RangeDet))$ function, which corresponds to \ac{MLE}. Hence, \ac{MAP} estimate converges to \ac{MLE} as $\NumOfSensingAntennas \to \infty$.
\end{proof}
\end{lemma}
\begin{lemma}
Let $u: \mathbb{R}^{\NumOfSensingAntennas\times1} \times \mathbb{R} \to \mathbb{R}$ be a one-to-one function. \ac{MLE} of $\RangeRV(\VectorSensingChsimple,x) \triangleq u(\VectorSensingChsimple x)= \sqrt[4]{\frac{x \RCS}{\VectorSensingChsimple}}$, where the \ac{PDF} $P_{\YSensingRVsl \mid \VectorSensingChsimple, X}$ is parameterised by $\VectorSensingChsimple$ (given $\TxsignalRVsl$). \ac{MLE} of $\RangeRV$ is:
\[
\Rshat = u(x \hat{H}_s),
\]
where $\hat{H}_s$ is \ac{MLE} of $\VectorSensingChsimple$, obtained by maximising $P_{\YSensingRVsl \mid \VectorSensingChsimple, X}$.
\label{invarianceMLE}
\begin{proof}
    See \cite[Theorem 7.2]{kay}. The invariance of the \ac{MLE} under-reparameterisation is shown by proving that the likelihood function and its maximiser remain unchanged by such transformations.
\end{proof}
\end{lemma}
\begin{theorem}
\ac{MLE} for estimating $h_s$ is simply the mean of the observations:
$\hat{h}_s
= \frac{1}{x}\frac{1}{\NumOfSensingAntennas} \sum_{j=1}^{\NumOfSensingAntennas} \boldsymbol{Y}_{s,j}$.
Thus \ac{MLE} for estimating $\RangeDet$ is,
$\Rshatdet = \begin{cases}
    \sqrt[4]{\frac{\RCS x}{\frac{1}{\NumOfSensingAntennas} \sum_{j=1}^{\NumOfSensingAntennas} \boldsymbol{Y}_{s,j}}} & \text{if } \frac{\RCS x}{\frac{1}{\NumOfSensingAntennas} \sum_{j=1}^{\NumOfSensingAntennas} \boldsymbol{Y}_{s,j}} \geq 0, \\
    \text{\ac{MLE} is not valid} & \text{otherwise}.
\end{cases}$
\label{thm:mle_hs}
\begin{proof}
$\hat{h}_s = \frac{1}{x} \min_{h_s \in \mathbb{R}^{+}} \| \YSensRVsldet - \VectorSensingChsimple(\RangeRV) \mathbf{1}_{\NumOfSensingAntennas} \|^2$ is an \ac{LS} problem with an analytical solution \cite{Boyd_Vandenberghe_2004}.
\end{proof}
\end{theorem}
\subsection{\texorpdfstring{\ac{BCRB}}{BCRB}-Based Approach: \texorpdfstring{\ac{OB}}{OB}}
\label{sec:bcrb}
\begin{theorem}
\label{theorem:bcrb}
The \ac{BCRB} for any unbiased estimator $\hat{\Rsh}$ of $\RangeRV$ with realisation $\RangeDet$ is given by:
$\text{BCRB}(x \mid \RangeDet) = \frac{1}{16 \NumOfSensingAntennas \RCS^2 x^2 \sigma^{-2}_\textrm{s} \RangeDet^{-8} + \lambda}$, where $\lambda$ is the rate parameter of the exponential prior distribution of $\RangeRV$.
\begin{proof}
See Appendix \ref{proof:bcrb}.
\end{proof}
\end{theorem}
\begin{lemma}
The \text{BCRB}$(\RangeDet \mid x)$ is asymptotically convex in $\RangeDet$ as either $\NumOfSensingAntennas$ or \ac{O-SNR} (or both) increase.
\label{lemma:cnvxity}
\begin{proof}
See Appendix \ref{lemma:convxity:proof}.
\end{proof}
\end{lemma}
\begin{remark}
The \ac{BCRB} is a valid lower bound for the \ac{MSE} of an estimator $\Rshat^\star$ only if it is unbiased \cite[Theorem 3.1]{kay}, which is ensured by a sufficiently large $\NumOfSensingAntennas$. Moreover, for large datasets, \ac{MLE} is asymptotically unbiased and achieves the \ac{BCRB} \cite[Theorem 11.3]{kay}.
\end{remark}
\begin{lemma}
\label{lemma:CRB}
Let $\mathbb{E}_{\RangeRV}[\text{BCRB}(\RangeRV \mid x)]$ denote the average sensing cost. This quantity is an asymptotic lower bound for the function $c(x)$ defined in \eqref{eq:c_x2}. Specifically, we have
\begin{equation}
\text{BCRB}(\mathbb{E}_{\RangeRV}[\RangeRV] \mid x) \leq \mathbb{E}_{\RangeRV}[\text{BCRB}(\RangeRV \mid x)] \leq c(x),
\label{eq:crb}
\end{equation}
The inequalities are asymptotic, supporting \ac{DRT} \cite{xiong2023torch} for state distribution in the regime of many sensing antennas.

\begin{proof}
The second inequality follows from \cite[Theorem 3.1 and 11.3]{kay}, and the first from Jensen's inequality \cite{Boyd_Vandenberghe_2004} and \Cref{lemma:cnvxity}, with equality when $P_{\RangeRV}(\RangeDet)$ is deterministic.
\end{proof}
\end{lemma}


\begin{corollary}
\label{cor1}
Based on \cref{lemma:CRB}, the expected \ac{BCRB} given $\RangeRV$, $\mathbb{E}_{\RangeRV}[\text{BCRB}(\RangeRV \mid x)]$, serves as an \ac{OB} for the optimal \ac{C-D} region in asymptotic, unbiased scenarios. When there is greater certainty (less randomness) in $\RangeRV$, sensing performance improves due to reduced variance; this allows the \ac{LB} in \eqref{eq:crb} to be achieved and minimises the Jensen gap. Conversely, prior state distributions with higher randomness tend to rely more on likelihood, which can potentially reduce bias around the mean in \eqref{eq:optimal est.}. Moreover, \Ac{MP} (the optimal estimator in \eqref{eq:optimal est.}) converges to a normal distribution as $\NumOfSensingAntennas$ increases, with variance decreasing by a factor of $\frac{1}{\NumOfSensingAntennas}$, regardless of the prior state distribution \cite{alma991028260009704706}. We refer to this phenomenon as the modified \ac{DRT}\footnote{This trade-off relates to the bias-variance tradeoff, prior-data balance, and prior vs. likelihood strength, which are discussed in statistical inference and machine learning literature \cite{alma991028260009704706}.}.
\end{corollary}

\begin{lemma}
The \ac{MAP} estimator has higher computational complexity due to the nested integrals required for computing $c(x)$, while the \ac{MLE} is less demanding. However, when the prior is known, the \ac{MAP} can outperform the \ac{MLE}. Their choice involves a trade-off between computational cost and performance, depending on the achievability of the \ac{BCRB}, system assumptions and required accuracy.
\begin{proof}
See Appendix \ref{appcomp1}.
\end{proof}
\end{lemma}
\begin{figure}[ht]
    \centering
    \begin{subfigure}{\linewidth}
        \centering
        \begin{tikzpicture}
            \begin{semilogyaxis}[
                yticklabel pos=left,
                ytick align=inside,
                grid=major,
                xlabel={Transmit Signal ($x$)},
                ylabel={Var($\hat{\RangeRV} \mid X=x$) [$m^2$]},
                legend style={
                    at={(0.53,1.01)}, 
                    anchor=north,
                    font=\footnotesize,
                    draw=none,
                    fill= none,
                },
                legend columns=2,
                ylabel style={font=\footnotesize},
                xlabel style={font=\footnotesize},
                ymin=5e-7, ymax=5e-1,
                xmin=0,xmax=30,
            ]
                \addplot+[
                    color=blue,
                    mark=square, 
                    smooth,
                    mark repeat=16, 
                ] table[x index=0, y index=6] {Figs/MultiAntennaSetup_lambda_0.5.dat};
                \addlegendentry{\ac{MLE}, $\lambda=0.5$}

                \addplot+[
                    color=blue,
                    mark=square,
                    style = dashed,
                    mark repeat=16, 
                ] table[x index=0, y index=6] {Figs/MultiAntennaSetup_lambda_0.3.dat};
                \addlegendentry{\ac{MLE}, $\lambda=0.3$}

                \addplot+[
                    color=red, 
                    smooth,
                    mark=asterisk,
                    mark repeat=16, 
                ] table[x index=0, y index=5] {Figs/MultiAntennaSetup_lambda_0.5.dat};
                \addlegendentry{\ac{MAP}, $\lambda=0.5$}

                \addplot+[
                    color=red, 
                    mark=asterisk,
                    style = dashed,
                    mark repeat=16, 
                ] table[x index=0, y index=5] {Figs/MultiAntennaSetup_lambda_0.3.dat};
                \addlegendentry{\ac{MAP}, $\lambda=0.3$}

                \addplot+[
                    color=black, 
                    mark=o,
                    mark repeat=16, 
                ] table[x index=0, y index=1] {Figs/MultiAntennaSetup_lambda_0.5.dat};
                \addlegendentry{\ac{BCRB}, $\lambda=0.5$}

                \addplot+[
                    color=black, 
                    mark=o, 
                    mark repeat=16, 
                ] table[x index=0, y index=1] {Figs/MultiAntennaSetup_lambda_0.3.dat};
                \addlegendentry{\ac{BCRB}, $\lambda=0.3$}
            \end{semilogyaxis}
        \end{tikzpicture}
        \caption{Variance of $\hat{\RangeRV}$.}
        \label{fig:variance}
    \end{subfigure}
    \begin{subfigure}{\linewidth}
        \centering
        \begin{tikzpicture}
            \begin{semilogyaxis}[
                yticklabel pos=left,
                ytick align=inside,
                grid=major,
                xlabel={Transmit Signal ($x$)},                ylabel={\ac{MSE}($\hat{\RangeRV} \mid X=x$) [$m^2$]},
                ylabel style={font=\footnotesize},
                xlabel style={font=\footnotesize},
                ymin=5e-7, ymax=5e-1,
                xmin=0, xmax=30,
                legend style={
                    at={(0.0,0.1)}, 
                    anchor=west,
                    font=\footnotesize,
                    draw=none,
                    fill=none,
                },
                legend columns=2,
            ]
                \addplot+[
                    smooth,
                    color=blue,
                    mark=square,
                    mark repeat=16, 
                ] table[x index=0, y index=9] {Figs/MultiAntennaSetup_lambda_0.5.dat};
                \addlegendentry{\ac{MLE},\; $\lambda=0.5$}

                \addplot+[
                    color=red,
                    mark=asterisk,
                    mark repeat=16, 
                ] table[x index=0, y index=8] {Figs/MultiAntennaSetup_lambda_0.5.dat};
                \addlegendentry{\ac{MAP},\; $\lambda=0.5$}

                \addplot+[
                    color=blue,
                    mark=square,
                    mark repeat=16, 
                    style=dashed,
                ] table[x index=0, y index=9] {Figs/MultiAntennaSetup_lambda_0.3.dat};
                \addlegendentry{\ac{MLE},\; $\lambda=0.3$}

                \addplot+[
                    color=red,
                    mark=asterisk,
                    mark repeat=16, 
                    style=dashed,
                ] table[x index=0, y index=8] {Figs/MultiAntennaSetup_lambda_0.3.dat};
                \addlegendentry{\ac{MAP},\; $\lambda=0.3$}
            \end{semilogyaxis}
        \end{tikzpicture}
        \caption{Mean Squared Error of $\hat{\RangeRV}$.}
        \label{fig:mse}
    \end{subfigure}
    \caption{Average \ac{BCRB}, and variance and \ac{MSE} of \ac{MAP} and \ac{MLE} for $\lambda=0.3$ and $0.5$ versus $x$ ($n_s=64$).}
    \label{fig:overallmultiantenna}
\end{figure}
\begin{table}[t]
\caption{Default Parameters.}
\label{tab:simulation_parameters}
\begin{tabularx}{\linewidth}{@{}l*{1}{l}l@{}}
\toprule
Parameter & Value & Description \\ 
\midrule
$\ScalarCommCh$ & 1 & Channel Coefficient \\ 
$\eta_0$ & 1 & Initial Learning Rate \\ 
$\gamma$ & 20 & Decay Rate \\ 
$\lambda$ & $0.3, 0.5 \, \text{m}^{-1}$ & Exponential Parameter \\ 
$\RCS$ & 1 & (Perfect) Reflectivity \\ 
$\SensingNoiseVar$, $\CommNoiseVar$ & 1 W & Noise Variances \\ 
$P$ & 10 W & Optical Power Budget \\ 
$q$ & 0.25 & quantisation Step \\ 
Noise Range & $[-5 \SensingNoiseVar, 5 \SensingNoiseVar]$ & Range \\ 
Number of Sensing Antennas & 1, 64 & Configuration \\ 
$x_{\lvert \Xset \rvert}$ & 30 & Last Mass Point in $\Xset$ \\ 
\bottomrule
\end{tabularx}
\end{table}
\section{Numerical Results}
\label{sec:num}
This section presents results based on \Cref{tab:simulation_parameters}.
For each estimator (\ac{MAP} or \ac{MLE}) and each $x \in \Xset$, we generate $N_r$ samples of $\RangeRV$ from $P_{\RangeRV} \sim \text{Exp}(\lambda)$ and $N_y$ samples of the sensing signal $\YSensRVsl$ from $P_{\YSensingRVsl \mid x, \RangeDet} \sim \mathcal{N} \left( \frac{\RCS x}{\RangeDet^4}, \SensingNoiseVar \right)$. The average sensing cost (\ac{MSE}) is approximated by $c(x) \approx \frac{1}{N_r} \sum_{\RangeDet^{[i]}} \frac{1}{N_y} \sum_{\YSensRVsldet^{[j]}} \left( \RangeDet^{[i]} - \Rshatdet(x, \YSensRVsldet^{[j]}) \right)^2$. The expectations of the variance and bias are computed similarly.

In \Cref{fig:overallmultiantenna}, as \ac{O-SNR} increases and $\lambda$ decreases from 0.5 to 0.3, \ac{MAP} and \ac{MLE} estimators converge, confirming the results from \Cref{maptoml}. In the single-antenna case (figures omitted for brevity), the performance difference between $\lambda = 0.3$ and $\lambda = 0.5$ is more noticeable, as $P_{\RangeRV}(\RangeDet)$ significantly influences performance with limited antennas. The increased bias in \ac{MLE} and \ac{MAP} for single antennas suggests potential violations of regularity conditions, rendering \ac{BCRB} an unreliable metric for sensing performance in this scenario. In contrast, \Cref{fig:overallmultiantenna} supports the modified \ac{DRT} defined in \cref{cor1}, for multi antenna setting, demonstrating that a more random distribution ($\lambda = 0.3$) enhances sensing performance, while a more deterministic distribution ($\lambda = 0.5$) degrades it.

\Cref{fig:cdf} shows the optimised \ac{CDF} for various modes in a multiple-antenna setting. The sensing-optimised input distribution, obtained via CVX \cite{cvx}, aligns with \Cref{lemm:sop}. We also present the high \ac{O-SNR} \ac{CDF} for the \ac{Com. Opt.} mode ($\text{Exp}(\frac{1}{E})$, from \cite{Moser}) and a common point ($t=10$) from the \ac{ISAC} optimised region. The similarity of \ac{ISAC}-optimised \ac{CDF}s across approaches confirms \Cref{theorem:bcrb,maptoml}, showing that stricter distortion constraints shift probability mass to $X > \epsilon$ and concentrate probabilities at specific points, validating \ac{DRT} of \ac{ISAC} in \ac{FSO} \ac{S-I} Gaussian channels with multiple antennas \cite{xiong2023torch}.
\begin{figure}[t]
\centering
\begin{subfigure}[t]{\linewidth}
\centering
    \begin{tikzpicture}
        \begin{semilogyaxis}[
            grid=major,
            xlabel={Transmit Signal ($x$)},
            ylabel={$P(X \leq x)$},
            legend columns=1, 
            legend style={at={(0.45,0.6)}, draw=none, fill=none, anchor=north west, font=\footnotesize},
            ylabel style={font=\footnotesize},
            xlabel style={font=\footnotesize},
            xmode=linear,
            ymode=linear,
            xmin=0,
            xmax=30,
            ymin=0,
            ymax=1,
        ]
            \addplot+[
                mark=ball,
                color=teal,
                mark color=teal,
                mark repeat=16,
                smooth,
                style=solid,
            ] table[x index=0, y index=4] {Figs/PDF_MultiAntennaSetup.dat};
            \addlegendentry{\ac{Com. Opt.} - \ac{CF}}

            \addplot+[
                color=red,
                mark=asterisk,
                smooth,
                mark repeat=16,
                style=solid,
            ] table[x index=0, y index=6] {Figs/PDF_MultiAntennaSetup.dat};
            \addlegendentry{\ac{Sens. Opt.} - \ac{MAP}}

            \addplot+[
                color=blue,
                smooth,
                mark=square,
                mark repeat=16,
                style=solid,
            ] table[x index=0, y index=8] {Figs/PDF_MultiAntennaSetup.dat};
            \addlegendentry{\ac{Sens. Opt.} - \ac{MLE}}

            \addplot+[
                color=black,
                mark=o,
                mark repeat=16,
                smooth,
                style=solid,
            ] table[x index=0, y index=10] {Figs/PDF_MultiAntennaSetup.dat};
            \addlegendentry{\ac{Sens. Opt.} - \ac{BCRB}}

            \addplot+[
                color=teal,
                style=dashed,
                mark=ball,
                smooth,
                mark size=2.3,
            ] table[x index=0, y index=2] {Figs/PDF_MultiAntennaSetup.dat};
            \addlegendentry{\ac{ISAC} - \ac{BCRB} - \ac{CF}}

            \addplot+[
                color=red,
                smooth,
                mark=asterisk,
                mark repeat=16,
            ] table[x index=0, y index=13] {Figs/PDF_MultiAntennaSetup.dat};
            \addlegendentry{\ac{ISAC} - \ac{MAP}}

            \addplot+[
                color=blue,
                mark=square,
                smooth,
                mark repeat=16,
            ] table[x index=0, y index=15] {Figs/PDF_MultiAntennaSetup.dat};
            \addlegendentry{\ac{ISAC} - \ac{MLE}}

            \addplot+[
                color=black,
                mark=o,
                style=dashed,
                smooth,
            ] table[x index=0, y index=11] {Figs/PDF_MultiAntennaSetup.dat};
            \addlegendentry{\ac{ISAC} - \ac{BCRB}}
        \end{semilogyaxis}
    \end{tikzpicture}
    \caption{}
    \label{fig:cdf}
\end{subfigure}
\begin{subfigure}[t]{\linewidth}
\centering
    \begin{tikzpicture}[spy using outlines={circle, magnification=3, size=1.7cm, connect spies}]
        \begin{semilogyaxis}[
            grid=major,
            xlabel={Distortion (\ac{MSE}) / \ac{BCRB} [$m^2$]},
            ylabel={Capacity [$\text{bps}/\text{Hz}$]},
            legend columns=1,
            legend style={at={(.5,.74)}, anchor=west, font=\footnotesize, draw=none, fill=none},
            ylabel style={font=\footnotesize},
            xlabel style={font=\footnotesize},
            xmode=log,
            ymode=linear,
            xmin=5e-6,
            xmax=1e-2,
            ymin=2e-1,
            ymax=6,
        ]
            \addplot+[
                color=teal,
                mark=ball,
                smooth,
                mark repeat=2,
            ] table[x index=0, y index=1] {Figs/CRCF_MultiAntennaSetup.dat};
            \addlegendentry{\ac{CF} - \ac{BCRB}}

            \addplot+[
                color=black,
                mark=o,
                smooth,
                mark repeat=2,
            ] table[x index=0, y index=1] {Figs/CRBCRB_MultiAntennaSetup.dat};
            \addlegendentry{\ac{BCRB}}

            \addplot+[
                color=red,
                mark=asterisk,
                smooth,
                mark repeat=2,
            ] table[x index=0, y index=1] {Figs/CRMAP_MultiAntennaSetup.dat};
            \addlegendentry{\ac{MAP}}

            \addplot+[
                color=blue,
                mark=square,
                smooth,
                mark repeat=2,
            ] table[x index=0, y index=1] {Figs/CRML_MultiAntennaSetup.dat};
            \addlegendentry{\ac{MLE}}

            \addplot+[magenta, no markers, style=dashed] coordinates {(5e-6,1.3506) (1e-2,1.3506)};
            \addlegendentry{\ac{Com. Opt.} (\ac{LB})}

            \addplot+[cyan, no markers, style=dashdotdotted] coordinates {(5e-6,4.7646) (1e-2,4.7646)};
            \addlegendentry{\ac{Com. Opt.} (\ac{UB})}

            \addplot+[purple, no markers, style=dashdotted] coordinates {(5e-6, 2.8157) (1e-2, 2.8157)};
            \addlegendentry{\ac{Com. Opt.} (\ac{BAA})}

            \addplot+[
                color=orange,
                mark=*,
                only marks,
                mark repeat=2
            ] coordinates {(5.7213e-06, 2e-1)};
\node at (axis cs: 5.7213e-06, 2e-1) 
    [pin=30: {\footnotesize \textbf{Sens. Opt.} - \textit{BCRB} - \textsc{CVX}}] {};

\addplot+[
    color=orange,
    mark=*,
    only marks,
    mark repeat=2
] 
coordinates {(1.1129e-05, 2e-1)};

\node at (axis cs: 1.1129e-05, 2e-1) 
    [pin={10: {\footnotesize \textbf{Sens. Opt.} - \textit{MAP} \& \textit{MLE} - \textsc{CVX}}}, 
    draw=orange, 
    line width=1.5pt] {}; 

\addplot+[
    color=orange,
    mark=*,
    only marks,
    mark repeat=2
] 
coordinates {(1.1590e-05, 2e-1)};
\spy [thick] on (1.8, 2.2)
             in node [left] at (1.8,3.4);
        \end{semilogyaxis}
    \end{tikzpicture}
    \caption{}
    \label{fig:rdregion}
\end{subfigure}
\caption{
(left) optimised \ac{CDF} for several modes,
(right) \ac{C-D} Region ($\NumOfSensingAntennas=64$).}
\label{fig:combined}
\end{figure}

\Cref{fig:rdregion} shows that \ac{BCRB}-based methods serve as an \ac{OB}, with \ac{MAP} and \ac{MLE} covering larger areas due to lower \ac{MSE}. In multi-antenna setups, \ac{BCRB} narrows the gap to \ac{MAP}/\ac{MLE}. Validation through \ac{Sens. Opt.} and \ac{Com. Opt.} modes show convergence of \ac{MAP} and \ac{MLE} as $\NumOfSensingAntennas$ increases, with the \ac{CF} region closely aligning with the \ac{BAA} region.
\section*{Conclusion}
In this paper, we revisited the performance of \ac{O-ISAC} from a \ac{C-D} perspective, developing practical \ac{MAP} and \ac{MLE} estimators for target distance that converge to the \ac{BCRB} as the number of sensing antennas increases. Our analysis established the \ac{R-CRB} as an asymptotic \ac{OB} for the optimal \ac{C-D} region and extended the \ac{DRT} for improved applicability in optical \ac{ISAC}. Additionally, we introduced iterative \ac{BAA}-type and memory-efficient algorithms for determining optimal input distributions, demonstrating that at high \ac{O-SNR}, the optimal input distribution belongs to the exponential family.

\section*{Acknowledgement}
This work is supported by the UK Department for Science, Innovation, and Technology under the Future Open Networks Research Challenge project TUDOR (Towards Ubiquitous 3D Open Resilient Network). The views expressed are those of the authors and do not necessarily represent the project.
\bibliographystyle{IEEEtran}
\bibliography{IEEEabrv,conf1_ref}
\appendices
\section{Proof of \texorpdfstring{\Cref{lemma:est}}{Lemma}}
\label{proof:lemma:est}
\begin{proof}
Following \cite[Appendix A]{mehrasajournal}, where a scalar feedback observation on the $i$-th channel (denoted as $z_i$ in \cite{mehrasajournal}, and $y_{s,i}$ here) is considered, we extend this to a vector of $\NumOfSensingAntennas$ independent observations.
$\Rshatdet^\star =
    \int_{\RangeDet \in \Rsset} P_{\RangeRV \mid \TxsignalRVsl, \YSensRVsldet}(\RangeDet \mid x, \YSensRVsldet) d(\RangeDet, \RangeDet^{\prime})$.
Then,
    $c(x) = \mathbb{E}_{\YSensingRV \mid \TxsignalRV} \left[ \mathbb{E}_{\instRangeRV \mid \TxsignalRV, \YSensingRV}[(\instRangeRV - \mu)^2] \right]$
\end{proof}

\section{\texorpdfstring{Numerical Methods for Solving \eqref{opt1}}{Numerical Methods for Solving}}
\label{app:numerical_methods}

To derive the \ac{C-D} region, we apply gradient descent for power allocation and \ac{BAA} (\BAAF) to optimise $s^{(i)}$ for each $t \in \mathcal{T}$, maximising \ac{MI} within the power budget. In high \ac{O-SNR} scenarios, the \ac{CFA} from \cref{lemma:kkt} (\CFAF) replaces \BAAF.
\begin{remark}
    In \Cref{alg.1}, $\Xset$ is quantised as:
    $\Xset_q = \{ (m-1)q \mid m = 1, \ldots, \frac{|\Xset|_q}{q} + 1 \}$,
    where $q$ is the quantisation step. The Gaussian noise $Z_c$ is quantised with step $q^2$, denoted as $\mathcal{Z}_{c,q^2}$. For small $q$, $\Xset_q \approx \Xset$ and $\mathcal{Z}_{c,q^2} \approx \mathcal{Z}_c$.
\end{remark}
\section{Proof of \texorpdfstring{\Cref{lemm:sop}}{Lemma}}
\begin{proof}
    Let us write the Lagrangian function for \eqref{opt3}:
    $J = \int_{x\in \Xset} c(x) P_X(x) + \eta_4 \big(\int_{x\in \Xset} x P_X(x) - P\big) + \eta_5 \big(\int_{x\in \Xset} P_X(x) - 1\big)$.
Differentiating $J$ with respect to $P_X(\TxsignalRVrealization)$ for each $i$ yields:
$\frac{\partial J}{\partial P_X(\TxsignalRVrealization)} = c(\TxsignalRVrealization) + \eta_4 \TxsignalRVrealization + \eta_5$.
Assuming $\eta_4 = 0$, we get $\eta_5 = -c(\TxsignalRVrealization)$ by complementary slackness, implying $c(\TxsignalRVrealization)$ must be constant, which is generally not the case. Hence, $\eta_4 \neq 0$. For optimal solution, complementary slackness requires:
$\int_{x \in \Xset} \TxsignalRVrealization P_X^{\text{\ac{Sens. Opt.}}}(\TxsignalRVrealization) = P$.
For a nonnegative random variable $\TxsignalRVsl$, $\mathbb{E}[X] \geq \min(X)$, with equality if and only if $\TxsignalRVsl$ is almost surely constant and equal to $\min(X)$ \cite[Theorem 2.6.1]{cover}.
So, $\min c(x) \leq \int_{x \in \Xset} c(x) P_X(x)$.
Thus, $P_X^{\text{\ac{Sens. Opt.}}}(x) = \delta(x - x^\star)$, where $x^\star \triangleq \arg \min_{x \leq P} c(x)$. 
\label{proof:sop}
\end{proof}
\section{Proof of \texorpdfstring{\Cref{theorem:bcrb}}{Theorem}}
\label{proof:bcrb}
\begin{proof}
For $\NumOfSensingAntennas$ samples, the log-likelihood $\mathcal{L} \triangleq \log p_{\YSensingRVsl \mid \TxsignalRVsl, \RangeRV}$ is: $\mathcal{L} = -\frac{\NumOfSensingAntennas}{2} \log(2 \pi N) - \frac{1}{2\SensingNoiseVar} \sum_{i=1}^{\NumOfSensingAntennas} (y_{s,i} - \RCS x \RangeDet^{-4})^2$.
The score function is: $\frac{\partial \mathcal{L}}{\partial \RangeDet} = -\frac{1}{\SensingNoiseVar} \sum_{i=1}^{\NumOfSensingAntennas} (y_{s,i} - \RCS x \RangeDet^{-4}) (4 \RCS x \RangeDet^{-3}) + \lambda$. The \ac{FIM} is: $I(\RangeDet) = \frac{16 \NumOfSensingAntennas \RCS^2 x^2 \RangeDet^{-8}}{\SensingNoiseVar} + \lambda$. Thus, the \ac{BCRB} is: $\text{BCRB}(x \mid \RangeDet) = \frac{1}{I(x,\RangeDet)} = \frac{1}{\frac{16 \NumOfSensingAntennas \RCS^2 x^2 \RangeDet^{-8}}{\SensingNoiseVar} + \lambda}$.
\end{proof}
\section{Proof of \texorpdfstring{\Cref{lemma:cnvxity}}{Lemma}}
\begin{proof}
\label{lemma:convxity:proof}
To prove the convexity of $f(\RangeDet) \triangleq \text{BCRB}(\RangeDet \mid x)$, start by simplifying the function:
$f(\RangeDet) = \frac{1}{A \RangeDet^{-8} + \lambda}, \quad \text{where } A = \frac{16 \NumOfSensingAntennas \RCS^2 x^2}{\SensingNoiseVar}$.
The first derivative is:
$f'(\RangeDet) = \frac{8 A \RangeDet^{-9}}{(A \RangeDet^{-8} + \lambda)^2}$.
The second derivative, using the quotient rule, is:
$f''(\RangeDet) = \frac{56 A^2 \RangeDet^{-8} - 72 A \lambda}{(A \RangeDet^{-8} + \lambda)^3}$.
For convexity, $f''(\RangeDet)$ should be non-negative. Since $A \RangeDet^{-8}$ dominates for large $A$, $f''(\RangeDet)$ is positive, proving the convexity of $f(\RangeDet)$.
\end{proof}
\section{\texorpdfstring{Complexity Comparison of \ac{BAA} and \ac{CFA}}{Complexity Comparison of BAA and CFA}}
\label{appcomp1}

This section compares the computational and memory complexities of the \ac{BAA} and \ac{CFA} algorithms for computing the \ac{C-D} region.
Both algorithms have a memory complexity of $\mathcal{O}(\lvert \Xset \rvert)$.
The \ac{BAA} algorithm has a computational complexity per iteration of $\mathcal(O)(\lvert\Xset\rvert \times \lvert\YCommRVsl\rvert)$, resulting in a total complexity of 
$O\big( |\mathcal{T}| \times N_{iter,BAA} \times \lvert\Xset\rvert \times \lvert\YCommRVsl\rvert \big)$, where $N_{iter,BAA}$ is the number of iterations required for convergence.
In contrast, the \ac{CFA} algorithm has a per-iteration complexity of $\mathcal(O)(\lvert\Xset\rvert)$, leading to a total complexity of 
$O\big( \lvert\mathcal{T}\rvert \times N_{iter,CFA} \times \lvert\Xset\rvert \big)$, where $N_{iter,CFA}$ is the number of iterations for convergence in the \ac{CFA} method.
\Cref{tab:complexity_comparison} summarizes the complexities:

\begin{table}[t]
    \centering
    \caption{Complexity Analysis of \ac{BAA} and \ac{CFA}}
    \label{tab:complexity_comparison}
    \resizebox{0.48\textwidth}{!}{%
    \begin{tabular}{|c|c|c|}
        \hline
        \textbf{Algorithm} & \textbf{Computational Complexity} & \textbf{Memory Complexity} \\
        \hline
        \ac{BAA} & $\mathcal{O}\big( |\mathcal{T}| \times N_{iter,BAA} \times |\Xset| \times |\YCommRVsl| \big)$ & $\mathcal{O}(|\Xset|)$ \\ \hline
        \ac{CFA} & $\mathcal{O}\big( |\mathcal{T}| \times N_{iter,CFA} \times |\Xset| \big)$ & $\mathcal{O}(|\Xset|)$ \\
        \hline
    \end{tabular}%
    }
\end{table}
\begin{corollary}
\Ac{BAA} provides more general \ac{C-D} region computation at a higher computational cost, whereas \ac{CFA} is less computationally complex. The choice between them depends on the \ac{O-SNR} for communication.
\end{corollary}
\section{\texorpdfstring{Complexity Comparison of \ac{MAP} and \ac{MLE}}{Complexity Comparison of MAP and MLE}}
\label{appcomp2}

This section compares the computational and memory complexities of the \ac{MAP} and \ac{MLE} for computing \eqref{eq:c_x2}.
The complexity for computing $c(x)$ with the \ac{MAP} estimator involves evaluating the nested integrals, resulting in a computational complexity of $\mathcal{O}(\lvert \Xset \rvert \times \lvert \YCommRVsl \rvert \times N_{iter,MAP})$, where $N_{iter,MAP}$ is the number of iterations required for convergence. The memory complexity is $\mathcal{O}(\lvert \Xset \rvert \times \lvert \YCommRVsl \rvert)$. For the \ac{ML} estimator, the computational complexity is given by $\mathcal{O}(\lvert \Xset \rvert \times N_{iter,ML})$, with a memory complexity of $\mathcal{O}(\lvert \Xset \rvert)$.
Additionally, the complexity of computing the $\text{BCRB}(x \mid \RangeDet)$, as referred to in \cref{theorem:bcrb}, is $\mathcal{O}(1)$ for both computational and memory complexities since it is a direct evaluation.
\Cref{tab:complexity_comparison_estimators} summarizes the complexities:

\begin{table}[t]
    \centering
    \caption{Complexity Analysis of Computing $c(x)$}
    \label{tab:complexity_comparison_estimators}
    \resizebox{0.48\textwidth}{!}{%
    \begin{tabular}{|c|c|c|}
        \hline
        \textbf{Estimator/\ac{BCRB}} & \textbf{Computational Complexity} & \textbf{Memory Complexity}\\
        \hline
        \ac{MAP} & $\mathcal{O}(\lvert \Xset \rvert \times \lvert \YCommRVsl \rvert \times N_{iter,MAP})$ & $\mathcal{O}(\lvert \Xset \rvert \times \lvert \YCommRVsl \rvert)$ \\ \hline
        \ac{ML} & $\mathcal{O}(\lvert \Xset \rvert \times N_{iter,ML})$ & $\mathcal{O}(\lvert \Xset \rvert)$ \\
        \hline
        \ac{BCRB} & $\mathcal{O}(1)$ & $\mathcal{O}(1)$ \\
        \hline
    \end{tabular}%
    }
\end{table}

\begin{corollary}
    The \ac{MAP} estimator has higher computational complexity due to the nested integrals required for computing $c(x)$, while the \ac{MLE} is less demanding. However, the \ac{MAP} can outperform the \ac{MLE} when the prior is known. The choice between them involves a trade-off between computational cost and performance, depending on the achievability of the \ac{BCRB}, system assumptions, and required accuracy.
\end{corollary}
\begin{algorithm*}[t]
    \caption{\ac{BAA}-type and \ac{CFA} Algorithms for the \ac{C-D} Region}
    \label{alg.1}
    \DontPrintSemicolon
        \KwInput{
        $\delta_{BA}$, $\delta_{B}$, $\Xset$, and $P$}
        \KwOutput{$\{D^{(j)}\}_{j=1}^{\lvert \mathcal{T} \rvert}$ and $\{\mathcal{I}^{(j)}\}_{j=1}^{\lvert \mathcal{I} \rvert}$}
        \tcp{\ac{BAA} Function for general}
        \Fn{\BAAF{$s^{(i)}, t, P_X$}}{
            \KwInit{
            $iter := 0$}
        \Repeat{$\left|\mathcal{I}_U - \mathcal{I}_L\right| < \delta_{BA}$}{
            $iter:=iter+1$

            \If{$iter \neq 1$}{
                $P_X^{\star} := \frac{P_X \odot \boldsymbol{g}}{2^{\mathcal{I}_L}}$
            }
            
            $g_k := \exp\Bigg( \sum_{y_c \in \YCommRVsl} P_{\mathsf{Y}_c \vert X}(y_c \vert x_k)$
            $\times \log_2\bigg( \frac{P_{\mathsf{Y}_c \vert X}(y_c \vert x_k)}{\sum_{y_c \in \YCommRVsl} P_{\mathsf{Y}_c \vert X}(y_c \vert x_k) P_X(x_k)} \bigg) \Bigg)$

            $\mathcal{I}_L := \log_2{(\sum_{x \in \Xset} g_k P_X(x_k))}$
            
            $\mathcal{I}_U := \log_2(\max_{k} g_k)$
            
            $iter := iter + 1$
        }
        \KwRet
        \newline
        $P_i = \mathcal{X}_q P_X,
        D_i = c(\mathcal{X}_q) P_X,
        \newline
        \mathcal{I}_i = s^{(i)} \times P_i + t \times D_i + \mathcal{I}_L$
        }
        \tcp{\ac{CFA} for high \ac{O-SNR}}
        \Fn{\CFAF{$s^{(i)}, t$}}{
        $G_X(x) = \exp{(-s^{(i)}\cdot x-t \cdot c(x))}$
        
        $P_X:=\frac{G_X(x)}{\sum_{j=0}^{\lvert \Xset\rvert}G_X(x_j)}$
        
        $P_i := \sum_{j=0}^{\lvert \Xset\rvert} x_j P_X(x_j)$
        
        $D_i := \sum_{j=0}^{\lvert \Xset\rvert} c(x_j) P_X(x_j)$
        
        \KwRet $P_i$, $D_i$, $P_X$, and $\mathcal{I}_i$
        }
        \tcp{main}
        \KwInit{$P_X := \frac{1}{\lvert \Xset \rvert}$, $j := 1$}
        \For{$t \in \mathcal{T}$}{
            $i := 0$
            $s^{(0)} := 0$
            
            \BAAF{$s^{(i)}, t,p_X$}
            
            \tcc{or
            \CFAF{$s^{(i)}, t$}}
            
            \If{$P_i<P$}{
            \textbf{Break}}
            \Else{
            $i := 1$
            $s^{(1)} := 1$
            
            \BAAF{$s^{(i)}, t,p_X$}
            
            \tcc{or
            \CFAF{$s^{(i)}, t$}}
            
            \Repeat{$\left|P_i - P\right| < \delta_{B}$}{
                $\eta_k = \frac{\eta_0}{1 + \gamma \cdot (0.1)^i}$
                $i := i + 1$
                $ds := \frac{P_i - P_{i-1}}{s^{(i)} - s^{(i-1)}}$
                
                $s^{(i)} := \max\{0, s - \eta_k \frac{P_i - P}{ds}\}$
                
            \BAAF{$s^{(i)}, t,p_X$}
            
            \tcc{or
            \CFAF{$s^{(i)}, t$}}
            }
            }
            $D^{(j)} := D_i$
            $\mathcal{I}^{(j)} := \mathcal{I}_i$
            $j := j + 1$
        }
\end{algorithm*}
\end{document}